\documentclass[12pt]{article}

\usepackage{amsmath}
\allowdisplaybreaks

\begin{document}

\baselineskip=18.5pt plus 0.2pt minus 0.1pt

\makeatletter
\@addtoreset{equation}{section}
\renewcommand{\theequation}{\thesection.\arabic{equation}}
\renewcommand{\thefootnote}{\fnsymbol{footnote}}

\newcommand{\ap}{\alpha'}
\newcommand{\p}{\partial}
\newcommand{\nn}{\nonumber}
\newcommand{\Me}{M_0}
\newcommand{\Mo}{M_1}
\newcommand{\ve}{\bm{v}_0}
\newcommand{\vo}{\bm{v}_1}
\newcommand{\Up}{U_{+}}
\newcommand{\Um}{U_{-}}
\newcommand{\go}{g_{\rm o}}
\newcommand{\calS}{{\cal S}}
\newcommand{\calQ}{{\cal Q}}
\newcommand{\calEc}{{\cal E}_{\rm c}}
\newcommand{\calNt}{{\cal N}_{\rm t}}
\newcommand{\calQB}{{\cal Q}_{\rm B}}
\newcommand{\calM}{{\cal M}}
\newcommand{\mt}{m_{\rm t}}
\newcommand{\Phit}{\Phi_{\rm t}}
\newcommand{\phit}{\phi_{\rm t}}
\newcommand{\vac}{\ket{0}}
\newcommand{\Psic}{\Psi_{\rm c}}
\newcommand{\phic}{\phi_{\rm c}}
\newcommand{\calNc}{{\cal N}_{\rm c}}
\newcommand{\calO}{{\cal O}}
\newcommand{\Pmatrix}[1]{\begin{pmatrix} #1 \end{pmatrix}}
\newcommand{\wt}[1]{\widetilde{#1}}
\newcommand{\wh}[1]{\widehat{#1}}
\newcommand{\bm}[1]{\boldsymbol{#1}}
\newcommand{\diag}{\mathop{\rm diag}}
\newcommand{\bra}[1]{\langle #1\vert}
\newcommand{\ket}[1]{\vert #1\rangle}
\newcommand{\braket}[2]{\langle #1\vert #2\rangle}
\newcommand{\bbbk}[4]{{}_1\langle #1|{}_2\langle #2|
                      {}_3\langle #3|#4\rangle_{123}}
\newcommand{\bum}{\bm{u}^{(-)}}
\newcommand{\bup}{\bm{u}^{(+)}}
\newcommand{\bupm}{\bm{u}^{(\pm)}}
\newcommand{\bump}{\bm{u}^{(\mp)}}
\newcommand{\UW}{W}
\newcommand{\calUp}{{\cal U}^{+}}
\newcommand{\calUm}{{\cal U}^{-}}
\newcommand{\Greg}{G_{\rm reg}}
\newcommand{\biginv}{{\cal R}}

\begin{titlepage}
\title{
\hfill\parbox{4cm}
{\normalsize KUNS-1743\\{\tt hep-th/0111034}}\\
\vspace{1cm}
{\bf Observables as Twist Anomaly in\\ Vacuum String Field Theory}
}
\author{
Hiroyuki {\sc Hata}
\thanks{{\tt hata@gauge.scphys.kyoto-u.ac.jp}}
\quad and \quad
Sanefumi {\sc Moriyama}
\thanks{{\tt moriyama@gauge.scphys.kyoto-u.ac.jp}}
\\[15pt]
{\it Department of Physics, Kyoto University, Kyoto 606-8502, Japan}
}
\date{\normalsize November, 2001}
\maketitle
\thispagestyle{empty}

\begin{abstract}
\normalsize
We reveal a novel mathematical structure in physical observables, the
mass of tachyon fluctuation mode and the energy density, associated
with a classical solution of vacuum string field theory constructed
previously [hep-th/0108150].
We find that they are expressed in terms of quantities which
apparently vanish identically due to twist even-odd degeneracy of
eigenvalues of a Neumann coefficient matrix defining the three-string
interactions.
However, they can give non-vanishing values because of the breakdown
of the degeneracy at the edge of the eigenvalue distribution.
We also present a general prescription of correctly simplifying the
expressions of these observables.
Numerical calculation of the energy density following our prescription
indicates that the present classical solution represents the
configuration of two D25-branes.
\end{abstract}

\end{titlepage}

\section{Introduction}

Vacuum string field theory (VSFT) \cite{VSFT1,VSFT2,VSFT3,VSFTR} has
been proposed as a string field theory expanded around the tachyon
vacuum \cite{Sen,Sen:1999mg,Sen:1999xm}.
The action of VSFT is simply given by that of ordinary cubic string
field theory (CSFT) with its BRST operator replaced by
$\calQ$ linear in the ghost coordinate:
\begin{equation}
\calQ=c_0+\sum_{n\ge 1}f_n\left(c_n+(-1)^nc_n^\dagger\right) .
\label{eq:calQ}
\end{equation}
Since the cohomology of $\calQ$ is trivial, VSFT expanded around
the trivial configuration $\Psi=0$ contains no physical open string
excitations at all. Therefore, VSFT around $\Psi=0$ is believed to
describe pure closed string theory though no direct proof for this
expectation has been given yet.

Another problem concerning VSFT is to show that it has classical
solutions representing D$p$-branes, in particular, D25-bane.
The energy density of these solutions relative to the trivial one must
be equal to the D$p$-brane tension.
The perturbation expansion around the D25-brane solution must
reproduce the ordinary bosonic open string theory.

The fact that the BRST operator $\calQ$ of VSFT consists purely of
the ghost coordinate makes it easier to solve its classical equation
of motion.
First, we can consider solutions factorized into the matter part and
the ghost one. Second, for this type solution, the equation of motion
for the matter part implies that it is a kind of projection operator
\cite{Half,GT1,KawaOku,GT2,FuruOku,Moeller:2001ap}.
Using these facts, the matter part of the solutions have been obtained
and the ratio of the energy densities of two D$p$-brane solutions
have been found to reproduce the expected tension ratio
\cite{VSFT1,VSFT2}.
In these analyses they assumed that the ghost part is common
among D$p$-brane solutions for all $p$, and hence it was unnecessary
to know the explicit form of the ghost part.

However, for studying whether the energy density of a D$p$-brane
solution itself, instead of the ratio, is equal to the correct one,
and whether the perturbation expansion around the solutions reproduce
the known open string theories, we have to obtain the solutions
including their ghost parts.
In \cite{HatKaw}, they constructed a translationally and Lorentz
invariant classical solution of VSFT including the ghost part, and
analyzed the fluctuation spectrum and the energy density of the
solution.
The mass of the tachyon fluctuation mode and the ratio of the energy
density to the D25-brane tension are given in closed forms using the
Neumann coefficients defining the three-string interactions.
They calculated these two quantities numerically using the level
truncation. Though the tachyon mass was correctly reproduced, the
calculation of the energy density did not give the expected value
of the D25-brane tension.

In this paper, we shall unmask beautiful mathematical structures of
physical observables such as the tachyon mass and the ratio of the
energy density to the D25-brane tension obtained in \cite{HatKaw}.
In \cite{HatKaw}, they gave the tachyon mass squared and the ratio in
the form $-\ln 2/G$ and $\pi^2/(3\ln^3 2)\exp(-6 H)$, respectively,
using $G$ and $H$ which are expressed in closed forms using
Neumann coefficients.
We find that both $G$ and $H$ are quantities which vanish identically
if we use the known identities among the Neumann coefficients.
On the other hand, numerical calculation of these quantities gave
non-vanishing results. We identify the origin of this paradox.
We argue that both $G$ and $H$ are quantities similar to the chiral
index of fermions \cite{Coleman} or the Witten index in supersymmetric
theories \cite{WittenIndex}. They almost vanish because of the
degeneracy of eigenvalues of a Neumann coefficient matrix due to
world-sheet twist transformation. Non-vanishing values of $G$ and $H$
come from the breakdown of degeneracy at the edge of the eigenvalue
distribution. Therefore we call this phenomenon twist anomaly.

As we mentioned above, it is dangerous to naively use the identities
among the Neumann coefficients to simplify the expressions of $G$ and
$H$.
We also present a general prescription for allowed deformation of
these quantities by taking into account the singularities at the
edge of the eigenvalue distribution.
By respecting the above lessons, we reexamine the energy density of
the classical solution. Our numerical calculation indicates that
the classical solution of \cite{HatKaw} represents two D25-branes if
there are no other subtle points.

The organization of the rest of this paper is as follows.
In sec.\ 2, after presenting the elements of the VSFT action, in
particular, the identities among the Neumann coefficients, we
summarize the classical solution presented in \cite{HatKaw}.
In sec.\ 3, we point out that the quantity $G$ giving the tachyon mass
vanishes if we naively use the identities,
and then resolve the paradox. We also present a general prescription
of allowed deformations for $G$ and $H$.
In sec.\ 4, we reexamine the energy density of the solution.
In the final section, we summarize the paper and discuss future
problems.

\section{VSFT and its classical solution}
In this section we shall first summarize basic elements of VSFT, in
particular, the Neumann coefficient algebra, and review its
translationally invariant classical solution given in \cite{HatKaw}.

\subsection{VSFT action}

The action of VSFT is given by \cite{VSFT1,VSFT2,VSFTR}
\begin{align}
\calS[\Psi]&=-K\biggl(\frac12\Psi\cdot\calQ\Psi
+\frac13\Psi\cdot(\Psi*\Psi)\biggr),
\label{eq:SV}
\end{align}
where the front factor $K$ is a constant. The BRST operator
$\calQ$ around the tachyon vacuum is given by a purely ghost form
(\ref{eq:calQ}), and satisfies the nilpotency and the Leibniz rule on
the $*$-product.
The three-string vertex defining the $*$-product is the same as in the
ordinary CSFT and is given in the momentum representation for the
center-of-mass $x^\mu$ as \cite{CST,Samuel,Ohta,GJI,GJII,IOS}
\begin{align}
\ket{V}_{123}&=\exp\left(
-\frac12\,\bm{A}^\dagger\,C\calM_3\bm{A}^\dagger
-\bm{A}^\dagger\bm{V} -\frac12 V_{00}(A_0)^2
+ (\mbox{ghost part})
\right)\ket{0}_{123}\nn\\
&\hspace*{5cm}\times(2\pi)^{26}\delta^{26}\left(p_1+p_2+p_3\right),
\label{eq:V}
\end{align}
with various quantities defined by
\begin{align}
&\bm{A}=\Pmatrix{a^{(1)}_n\\a^{(2)}_n\\a^{(3)}_n},
\quad
A_0=\Pmatrix{a^{(1)}_0\\a^{(2)}_0\\a^{(3)}_0},
\quad
\calM_3 =\Pmatrix{M_0&M_+&M_-\\ M_-&M_0&M_+\\ M_+&M_-&M_0},
\nn\\
&\bm{V} =
\Pmatrix{\bm{v}_0&\bm{v}_+&\bm{v}_-\\
         \bm{v}_-&\bm{v}_0&\bm{v}_+\\
         \bm{v}_+&\bm{v}_-&\bm{v}_0}\!A_0,
\quad
V_{00}=\frac12\ln\left(\frac{3^3}{2^4}\right).
\label{eq:newquantities}
\end{align}
The boldface letters, $\bm{A}$ and $\bm{V}$, are the vectors in the
level number space.
The matter oscillator $a^{(r)\mu}_n$ ($n\ge 1$) satisfies the
commutation relation,
\begin{equation}
[a^{(r)\mu}_n,a^{(s)\nu\dagger}_m]
=\eta^{\mu\nu}\delta_{nm}\delta^{rs},
\label{CR}
\end{equation}
and $a^{(r)}_0$ is related to the center-of-mass momentum of the
string $r$, $p_r=-i\p/\p x_r$, by $a^{(r)}_0=\sqrt{2}\,p_r$
(we are adopting the convention of $\ap=1$).
The real and symmetric matrices $(M_0)_{nm}$ and $(M_\pm)_{nm}$ and
the real vectors $(\bm{v}_0)_n$ and $(\bm{v}_\pm)_n$ in the level
number space are essentially the Neumann coefficient matrices
(see \cite{HatKaw} for their relation to the conventional Neumann
coefficients).
Finally, $C$ is the twist matrix defined by
\begin{equation}
C_{nm}=(-1)^n\delta_{nm},\qquad(n,m\ge 1).
\label{eq:C}
\end{equation}
It should be noted that the inner products in the exponent of
(\ref{eq:V}) are those in both the infinite dimensional level number
space and the three dimensional space of the strings $r=1,2,3$ (we
have omitted the transpose symbol for the vectors multiplying from the
left).

In the rest of this subsection, we shall summarize the algebras of
the Neumann coefficients $M_\alpha$ and $\bm{v}_\alpha$
($\alpha=0,\pm$).
First, the twist transformation property of the vertex,
\begin{equation}
\Omega_1\Omega_2\Omega_3\ket{V}_{123}=\ket{V}_{321},
\label{eq:OmegaOmegaOmegaV=V}
\end{equation}
is translated to the following for the Neumann coefficients:
\begin{equation}
CM_0C=M_0,\quad CM_\pm C=M_\mp,\quad
C\bm{v}_0=\bm{v}_0,\quad C\bm{v}_\pm=\bm{v}_\mp.
\label{eq:CVC=V}
\end{equation}
Here, $\Omega_r$ is the twist operator on the Fock space of the string
$r$:
\begin{equation}
\Omega\,a_n\Omega^{-1}=C_{nm}a_m,
\qquad\Omega\vac=\vac .
\label{eq:Omega}
\end{equation}
Next, they enjoy the following linear relations:
\begin{equation}
M_0+M_++M_-=1,\quad\bm{v}_0+\bm{v}_++\bm{v}_-=0.
\label{Linear}
\end{equation}
Therefore, let us take $(\Me, \Mo)$ and $(\ve, \vo)$
with $\Mo$ and $\vo$ defined by
\begin{equation}
\Mo=M_+-M_-,\qquad\vo=\bm{v}_+-\bm{v}_-,
\end{equation}
as independent quantities. Note that $\Me$ and $\ve$ are twist-even,
while $\Mo$ and $\vo$ are twist-odd:
\begin{equation}
C\Mo C=-\Mo, \quad C\vo=-\vo.
\end{equation}
Then, $\Me$, $\Mo$, $\ve$ and $\vo$ are known to satisfy the following
non-linear identities \cite{GJI,GJII,Kis}:
\begin{align}
&[\Me,\Mo]=0,\label{[M,M]}\\
&\Mo^2=(1-\Me)(1+3\Me),\label{MoOfMe}\\
&3(1-\Me)\ve+\Mo\vo=0,\label{Mv}\\
&3\Mo\ve +(1+ 3\Me)\vo=0,\\
&\frac94\,\ve^2 + \frac34\vo^2=2\, V_{00}\label{vv}.
\end{align}

\subsection{Classical solution of VSFT}
Now let us proceed to reviewing the construction of a translationally
and Lorentz invariant classical solution $\Psic$ to the equation of
motion of the VSFT action (\ref{eq:SV}) \cite{KP,VSFT2,HatKaw}:
\begin{equation}
\calQ\Psic+\Psic*\Psic=0 .
\label{eq:eqmot}
\end{equation}
The solution is expected to represent a space-time filling D25-brane.

We adopt the Siegel gauge for $\Psic$, $\ket{\Psic}=b_0\ket{\phic}$,
and assume the following form for $\ket{\phic}$:
\begin{equation}
\ket{\phic}=\calNc\exp\biggl(
-\frac12\sum_{n,m\ge 1}a^\dagger_n(CT)_{nm}a^\dagger_m
+\sum_{n,m\ge 1}c^\dagger_n(C\wt{T})_{nm}b^\dagger_m\biggr)\vac,
\label{eq:phic}
\end{equation}
where $T_{nm}$ and $\wt{T}_{nm}$ are unknown real matrices and
$\calNc$ is the normalization factor. We assume further that the state
$\ket{\phic}$ is twist invariant, $\Omega\ket{\phic}=\ket{\phic}$, and
hence $T_{nm}$ and $\wt{T}_{nm}$ satisfy the matrix equations
\begin{equation}
CTC=T,\qquad C\wt{T}C=\wt{T}.
\label{eq:twistinv}
\end{equation}
Then, $\Psic$ solves the equation of motion (\ref{eq:eqmot}) provided
the following two conditions are satisfied:
\begin{itemize}
\item
$T$ and $\wt{T}$ satisfy \begin{equation}
T=M_0+(M_+,M_-)(1-T\calM)^{-1}\,T\Pmatrix{M_-\\ M_+},
\label{eq:eqforT}
\end{equation}
with
\begin{equation}
\calM=\Pmatrix{M_0&M_+\\ M_-&M_0},
\label{eq:calM}
\end{equation}
and the same one with all the matrices replaced by
the tilded ones for the ghost oscillators, respectively.
The matrix $T$ on the RHS of (\ref{eq:eqforT}) should read
$\diag(T,T)$.

\item The normalization factor $\calNc$ is given by
\begin{equation}
\calNc=-\left[\det(1-T\calM)\right]^{13}
[\det(1-\wt{T}\wt{\calM})]^{-1}.
\label{eq:calNc}
\end{equation}
\end{itemize}
The arbitrary coefficient $f_n$ in the BRST operator $\calQ$
(\ref{eq:calQ}) is not a quantity which is given apriori, but rather it
is uniquely fixed by the requirement that the there exist a Siegel
gauge solution assumed above.
See \cite{HatKaw} for details.\footnote{
A concise expression of the coefficient $f_n$ is given in \cite{Kis}.}

Eq.\ (\ref{eq:eqforT}) for $T$ has been solved in \cite{KP,VSFT2}, and
we shall summarize the points in obtaining the solution.
Let us assume that $T$ commutes with the matrices $M_\alpha$:
\begin{equation}
[T,M_\alpha]=0,\qquad(\alpha=0,\pm).
\label{eq:[T,M]=0}
\end{equation}
Using the formulas (\ref{[M,M]}) and (\ref{MoOfMe}) for
$M_\alpha$ and, in particular,
\begin{equation}
\left(1-T\calM\right)^{-1}=\left(1-2M_0T+M_0T^2\right)^{-1}
\Pmatrix{1-TM_0&TM_+\\ TM_-&1-TM_0},
\label{eq:(1-TM)^-1}
\end{equation}
eq.\ (\ref{eq:eqforT}) is reduced to
\begin{equation}
(T-1)\Bigl(M_0T^2-(1+M_0)T+M_0\Bigr)=0.
\label{eq:(T-1)()=0}
\end{equation}
We do not adopt the solution $T=1$ which corresponds to the identity
state, and take a solution to
\begin{equation}
M_0T^2-(1+M_0)T+M_0=0.
\label{eq:M0T^2-(1+M0)T+M0=0}
\end{equation}
As a solution to (\ref{eq:M0T^2-(1+M0)T+M0=0}) we take
\begin{equation}
T=\frac{1}{2 M_0}\left(1+M_0-\sqrt{(1-M_0)(1+3M_0)}\right).
\label{eq:T-}
\end{equation}
The matrix square root in (\ref{eq:T-}) is defined as the
positive branch of the square root of the eigenvalues of the symmetric
matrix $(1-M_0)(1+3M_0)$.\footnote{See sec.\ 3 for the eigenvalue
distribution of $M_0$.}
It has been claimed by numerical comparison that the matter part of
the solution given by the present $T$ is equal to the sliver state
constructed by CFT arguments \cite{TSU,VSFT2}.

\section{Tachyon mass as twist anomaly}

In this section, we shall reexamine the mass of the tachyon
fluctuation mode around $\Psic$ obtained in \cite{HatKaw}. We shall
find that it has an interesting interpretation as a kind of
anomaly. For this purpose, we shall first summarize the construction
of the tachyon wave function.

\subsection{Tachyon fluctuation mode}
In \cite{HatKaw} fluctuation spectrum around the classical solution
$\Psic$ was also studied. Expanding the original string field
$\Psi$ in VSFT as
\begin{equation}
\Psi=\Psic+\Phi,
\label{eq:Psi=Psic+Phi}
\end{equation}
with $\Phi$ being the fluctuation, the VSFT action (\ref{eq:SV}) is
expressed as
\begin{equation}
\calS[\Psi]=\calS[\Psic]-K\biggl(\frac12\Phi\cdot\calQB\Phi
+\frac13\Phi\cdot(\Phi*\Phi)\biggr),
\label{eq:SV=SVc+S}
\end{equation}
where $\calQB$ is defined by
\begin{equation}
\calQB\Phi=\calQ\,\Phi + \Psic *\Phi +\Phi *\Psic .
\label{eq:calQB}
\end{equation}
The new BRST operator $\calQB$ also satisfies the nilpotency and the
Leibniz rule on the $*$-product.

We shall recapitulate the construction of tachyon wave function
$\Phit$ given in \cite{HatKaw}.
It is a scalar solution to
\begin{equation}
\calQB \Phit=0 ,
\label{eq:calQBPhit=0}
\end{equation}
and carries center-of-mass momentum $p^2=1$.
We take again the Siegel gauge for $\Phit$,
$\ket{\Phit}=b_0\ket{\phit}$, and assume the following form for
$\ket{\phit}$:
\begin{align}
\ket{\phit}&=\frac{\calNt}{\calNc}
\exp\biggl(-\sum_{n\ge 1}t_n a^\dagger_n a_0\biggr)\ket{\phic}.
\label{eq:phit}
\end{align}
Though not written explicitly, $\ket{\phit}$ carries
non-vanishing momentum in contrast with $\ket{\phic}$ which is
translationally invariant.
Since $\ket{\phit}$ is twist invariant, the vector $\bm{t}$
satisfies
\begin{equation}
C\bm{t}=\bm{t}.
\label{eq:Ct=t}
\end{equation}
The normalization factor $\calNt$ for $\ket{\phit}$ will be fixed later.
Then, the wave equation (\ref{eq:calQBPhit=0}) holds for the present
$\Phit$ if the vector $\bm{t}$ satisfies
\begin{align}
&\bm{t}=\bm{v}_0-\bm{v}_++(M_+,M_-)(1-T\calM)^{-1}T
\Pmatrix{\bm{v}_+-\bm{v}_-\\ \bm{v}_--\bm{v}_0}
+(M_+,M_-)(1-T\calM)^{-1}\Pmatrix{0\\ \bm{t}},
\label{eq:u}
\end{align}
and the center-of-mass momentum $p_\mu$ is subject to
$p^2=-\mt^2$ with the tachyon mass $\mt$ given by
\begin{equation}
-\mt^2=\frac{\ln 2}{G},
\label{eq:p^2fromG}
\end{equation}
in terms of $G$ defined by
\begin{align}
&G=2V_{00}+(\bm{v}_--\bm{v}_+,\bm{v}_+-\bm{v}_0)(1-T\calM)^{-1}T
\Pmatrix{\bm{v}_+-\bm{v}_-\\ \bm{v}_--\bm{v}_0}\nn\\
&\hspace{2cm}
+2(\bm{v}_--\bm{v}_+,\bm{v}_+-\bm{v}_0)
(1-T\calM)^{-1}\Pmatrix{0\\\bm{t}}
+(0,\bm{t})\calM(1-T\calM)^{-1}\Pmatrix{0\\ \bm{t}}.
\label{eq:G}
\end{align}

We have to solve (\ref{eq:u}) for the vector $\bm{t}$ to check whether
$\mt^2$ given by (\ref{eq:p^2fromG}) really reproduces the correct
value of the tachyon mass, $-\mt^2=1$.
First, eq.\ (\ref{eq:u}) was solved to give
\begin{equation}
\bm{t}=3(1+T)(1+3M_0)^{-1}\ve.
\label{eq:t}
\end{equation}
Putting this solution into (\ref{eq:G}) the following expression for
$G$ was obtained:
\begin{equation}
G=\frac94\ve\frac{1}{1+3M_0}\wh{G}\ve
-\frac14\vo\frac{1}{1-M_0}\wh{G}\vo,
\label{G}
\end{equation}
with
\begin{equation}
\wh{G}=\frac{-(1+3M_0)^2(1-2M_0)+(1-3M_0)\sqrt{(1-M_0)(1+3M_0)}}
{2M_0(1+3M_0)}.
\label{Ghat}
\end{equation}
The square root in (\ref{Ghat}) came from that in $T$ (\ref{eq:T-})
and hence the same branch should be taken.

\subsection{Reexamination of the tachyon mass}

In \cite{HatKaw} they used the level truncation to evaluate
the quantity $G$ numerically and found that it reproduces to high
precision the expected value $G=\ln 2$.
However, by using the formulas (\ref{[M,M]}), (\ref{MoOfMe}) and
(\ref{Mv}) in the expression (\ref{G}), we can show that $G$ vanishes
identically!
In fact, plugging
\begin{equation}
\ve=-\frac{1}{3(1-\Me)}\Mo\vo,
\label{eq:vebyvo}
\end{equation}
obtained from (\ref{Mv}) into the first term on the RHS of (\ref{G})
and using the commutativity (\ref{[M,M]}) and the formula
(\ref{MoOfMe}), we find that the two terms on the RHS cancel each
other. Note that this cancellation occurs for any $\wh{G}$ commutative
with $\Me$ and $\Mo$.
Numerical analysis of $G$ gives a finite and non-vanishing result,
while we can analytically show that the same quantity vanishes
identically. Where does this paradox come from?

As a preparation for understanding the origin of the paradox,
we shall mention the eigenvalue distribution of $\Me$.
We have evaluated the eigenvalues of $M_0$ numerically by the level
truncation, namely, by cutting off the size of the infinite
dimensional matrix to a finite $L\times L$ one.
The result is as follows. First, all the eigenvalues of $\Me$ are
negative and most of them are close to zero. Next, table \ref{tab:M0ev}
shows the three smallest eigenvalues for various values of the cutoff
$L$. The smallest eigenvalue $\lambda_1$ at $L=\infty$ is the result
of extrapolation by a fitting function of the form
$\sum_{k=0}^{10} c_k(\ln L)^{-k}$.
{}From this analysis and taking the arguments below into account
in advance, it is very likely that the smallest eigenvalue of
$\Me$ converges to $-1/3$ in the limit $L\to\infty$.
The eigenvalue distribution of $\Me$ would be continuous in the range
$(-1/3,0)$.
\begin{table}[htbp]
\begin{center}
\begin{tabular}[b]{|r|c|c|c|}
\hline
$L$~~ & $\lambda_1$ & $\lambda_2$ & $\lambda_3$ \\
\hline\hline
$50$ & $-0.28120$ & $-0.17637$ & $-0.08862$ \\
$100$ & $-0.28963$ & $-0.19607$ & $-0.11006$ \\
$150$ & $-0.29367$ & $-0.20618$ & $-0.12190$ \\
$200$ & $-0.29621$ & $-0.21277$ & $-0.12994$ \\
$250$ & $-0.29802$ & $-0.21758$ & $-0.13598$ \\
$300$ & $-0.29940$ & $-0.22133$ & $-0.14077$ \\
$400$ & $-0.30143$ & $-0.22691$ & $-0.14808$ \\
$500$ & $-0.30288$ & $-0.23098$ & $-0.15354$ \\
$600$ & $-0.30399$ & $-0.23415$ & $-0.15787$ \\
$800$ & $-0.30563$ & $-0.23888$ & $-0.16446$ \\
$1000$ & $-0.30681$ & $-0.24234$ & $-0.16937$ \\
$\infty$ &$-0.33342$ & &\\
\hline
\end{tabular}
\caption{
The three smallest eigenvalues of $\Me$ for various values of the
cutoff $L$.
}
\label{tab:M0ev}
\end{center}
\end{table}

Let us return to the paradox.
To identify the origin of the paradox and interpret it
as twist anomaly, let us carry out
the following formal argument by assuming that the eigenvalue
distribution of $\Me$ is in the range $(-1/3,0)$ and for simplicity
that it is discrete.
Then, note first that the eigenvalues $\lambda$ of $\Me$ are
two-fold degenerate except at $\lambda=-1/3$.
In fact, as seen by using the formulas (\ref{[M,M]}) and (\ref{MoOfMe}),
the twist-even and odd eigenvectors $\bupm_\lambda$ of $\Me$
corresponding to the eigenvalue $\lambda$ and satisfying
\begin{equation}
\Me\bupm_\lambda=\lambda\bupm_\lambda,\quad
C\bupm_\lambda=\pm\bupm_\lambda,\quad
\bm{u}^{(s)}_\lambda\!\cdot\bm{u}^{(s')}_{\lambda'}
=\delta_{s,s'}\delta_{\lambda,\lambda'},
\end{equation}
are related by
\begin{equation}
\bupm_\lambda=\frac{1}{\sqrt{(1-\lambda)(1+3\lambda)}}
\Mo\bump_\lambda .
\end{equation}
However, for $\lambda=-1/3$ degeneracy does not occur in general since
we have $\Mo\bupm_{\lambda=-1/3}=0$ due to (\ref{MoOfMe}).
Because the eigenvector corresponding to the lowest eigenvalue
$\lambda_1$ in table \ref{tab:M0ev} is twist-odd,\footnote{
Numerical evaluation of the eigenvectors of $\Me$ shows that they are
alternatively twist-even and odd as the eigenvalue increases.
}
we assume that the eigenvector corresponding to
$\lambda=-1/3$ is twist-odd and there is no corresponding twist-even
eigenvector.
Then, we expand $\vo$ and $\ve$ in terms of $\{\bum_\lambda\}$
and $\{\bup_\lambda\}$, respectively:
\begin{align}
\vo=\sum_\lambda A_\lambda \bum_\lambda ,
\quad
\ve=-\frac13\sum_{\lambda\ne -1/3} \sqrt{\frac{1+3\lambda}{1-\lambda}}
A_\lambda \bup_\lambda ,
\end{align}
where the coefficients in $\ve$ has been determined
by using (\ref{Mv}).
Plugging these expansions into $G$ (\ref{G}), we obtain
\begin{equation}
G=\biggl(\sum_{\lambda\ne -1/3}-\sum_{\lambda}\biggr)
\frac{\wh{G}(\lambda)}{4(1-\lambda)} A_\lambda^2 ,
\label{eq:Gisalmost0}
\end{equation}
where $\wh{G}(\lambda)$ is given by (\ref{Ghat}) with the matrix $\Me$
replaced by its eigenvalue $\lambda$.

Eq.\ (\ref{eq:Gisalmost0}) implies that $G$ is a quantity similar to
the chiral index of fermions \cite{Coleman} or the Witten
index in supersymmetric theories \cite{WittenIndex}. It almost
vanishes due to cancellation between twist-odd and even
contributions. However, owing to the mismatch at $\lambda=-1/3$,
$G$ can be non-vanishing.
Therefore, we call such phenomenon twist anomaly.
So far we have assumed that the eigenvalues of $\Me$ are
discrete. However, the actual eigenvalue distribution of $\Me$ would
be continuous near $\lambda=-1/3$, and a refinement is of course
necessary for (\ref{eq:Gisalmost0}).

Note that we do not have non-vanishing result for
(\ref{eq:Gisalmost0}) for all $\wh{G}$ other than (\ref{Ghat}).
Let us consider the following example.
By using the formulas (\ref{[M,M]})---(\ref{Mv}), the LHS of
(\ref{vv}) is expressed both as $9\ve(1+3\Me)^{-1}\ve$ and as
$\vo(1-\Me)^{-1}\vo$, which are respectively the first term and the
negative of the second term on the RHS of (\ref{G}) with $\wh{G}$
replaced with $4$. Since both of these two expressions should give the
finite value $\ln(3^3/2^4)$ as seen from (\ref{vv}), the RHS of
(\ref{G}) with $\wh{G}=4$ should vanish without any ambiguity.
In fact, we have calculated numerically these two quantities by the
level truncation to confirm that they both give values close to
$\ln(3^3/2^4)$.

This observation for $\wh{G}=4$ implies that we need a
singularity in $\wh{G}$ at $\Me=-1/3$ which would make divergent each
of the two terms on the RHS of (\ref{G}) or (\ref{eq:Gisalmost0})
and hence amplify the effect of the breakdown of the degeneracy.
In fact, for the genuine $\wh{G}$ given by (\ref{Ghat}), we have
\begin{equation}
\wh{G}= -\frac{2\sqrt{3}}{\sqrt{1+3\Me}}
-\frac{3\sqrt{3}}{4}\sqrt{1+3\Me}+\cdots ,
\label{eq:LaurentG}
\end{equation}
around $\Me=-1/3$.

\renewcommand{\arraystretch}{1.3}
\begin{table}[htbp]
\begin{center}
\begin{tabular}[b]{|c|c|c|c|c|}
\hline
$1/\sqrt{1+3M_0}$&$\Mo$&$\ve$&$\vo$&$\bm{t}$\\
\hline\hline
$1$&$-1$&$0$&$1$&$1$\\
\hline
\end{tabular}
\caption{Degree of singularity for various quantities.}
\label{tab:degree}
\end{center}
\end{table}
\renewcommand{\arraystretch}{1}

Now, we shall explain how to deform the expressions of $G$ and other
quantities interpretable as twist anomaly by respecting their
singularity at $\Me=-1/3$.\footnote{
Our deformation rule is an empirical one obtained from numerical works 
and still lacks rigorous mathematical justification.
}
Of course, calculating these quantities using their original
expressions like (\ref{eq:G}) is quite all right.
However, by deformations explained below, we can obtain simpler
expressions giving the same numerical result. At the same time, the
following arguments would be instructive for understanding the
phenomenon of twist anomaly.
Consider any quantity $F$ given as
\begin{equation}
F=\sum_{\alpha,\beta=0,1}f_{\alpha\beta}\,
\bm{v}_\alpha\calO_{\alpha\beta}(M_0,M_1)\bm{v}_\beta ,
\label{eq:F}
\end{equation}
where $f_{\alpha\beta}$ is a scalar coefficient and
$\calO_{\alpha\beta}(M_0,M_1)$ is a matrix valued function of $M_0$ and
$M_1$. Let this $F$ vanish like $G$ if we use naively the non-linear
relations (\ref{[M,M]})---(\ref{Mv}).
We are allowed, if we wish, to use (\ref{eq:vebyvo}) to express $\ve$
in terms of $\vo$ in (\ref{eq:F}). However, we must keep the original
ordering among $\Me$ and $\Mo$.
We can make the following simplifications in calculating $F$.
For this purpose, we assign the ``degree of singularity'' at
$\Me=-1/3$ for various quantities as given table \ref{tab:degree}.
Note that this assignment is compatible with all the non-linear
relations (\ref{[M,M]})---(\ref{vv}) and the definition
(\ref{eq:t}) of $\bm{t}$.
Then, we Laurent-expand $\calO_{\alpha\beta}(M_0,M_1)$ with respect to
$\Me$ around $\Me=-1/3$ by keeping the ordering among the matrices,
and count the degree of singularity of each term contributing to
(\ref{eq:F}) by summing the degree of the constituents.
If the degree of singularity of a term is less than three, we are
allowed to freely use all the non-linear relations
(\ref{[M,M]})---(\ref{Mv}) to simplify this term. However, if the
degree of singularity is equal to three, we must treat this term as
it stands.

For example, following the above rule, $G$ of (\ref{G}) is expressed
as
\begin{equation}
G= -\frac{9\sqrt{3}}{32}\,\vo\biggl(
\Mo\frac{1}{(1+3\Me)^{3/2}}\Mo-\frac43\frac{1}{\sqrt{1+3\Me}}
\biggr)\vo+\Greg ,
\label{eq:Gdeform}
\end{equation}
where $\Greg$ represents the term with degree less than three.
Since the whole of (\ref{eq:Gdeform}) vanishes by
naively using the non-linear relations,
we can identify $\Greg$ without explicit calculation starting from the
original $G$.
Namely, $\Greg$ is equal to the negative of the first term of
(\ref{eq:Gdeform}) calculated by naively using the non-linear
relations,
\begin{equation}
\Greg=-\frac{3\sqrt{3}}{32}\,\vo\sqrt{1+3\Me}\vo ,
\end{equation}
whose degree of singularity is equal to one.\footnote{
Another simple expression of $G$ is obtained by replacing $\wh{G}$ in
(\ref{G}) with the first singular term of (\ref{eq:LaurentG}).
}
By numerical calculation we find that (\ref{eq:Gdeform}) reproduces
a value close to $\ln 2$.

We have emphasized above that it is in general dangerous to freely use
the non-linear relations (\ref{[M,M]})---(\ref{Mv}). However, we used
them in solving (\ref{eq:eqforT}) for $T$ and (\ref{eq:u}) for
$\bm{t}$. We also used non-linear relations in obtaining (\ref{G})
from (\ref{eq:G}). In the rest of this section, we shall discuss
the validity of these manipulations.

First, let us reexamine the deformation from (\ref{eq:G}) to
(\ref{G}). A possible problem in this deformation is the use of
(\ref{eq:(1-TM)^-1}) which has been obtained by use of non-linear
relations.
Here we shall consider $(1-T\calM)^{-1}$ without
using the non-linear relations. To perform it, let us split $1-T\calM$
into the twist-even part $S$ and the twist-odd part $A$ as
$1-T\calM=S+A$ with
\begin{equation}
S=\Pmatrix{1-T\Me&-T(1-\Me)/2\\ -T(1-\Me)/2&1-T\Me},\qquad
A=\Pmatrix{&-T\Mo/2\\ T\Mo/2&}.
\end{equation}
Then we have
\begin{equation}
\frac{1}{S+A}=\frac1S-\frac1S A\frac1S
+\frac1S A\frac1S A\frac1S-\cdots .
\label{eq:(S+A)^-1}
\end{equation}
The inverse of $S$ is simply given as follows because
it is given purely in terms of $M_0$:
\begin{equation}
\frac{1}{S}=\frac{1}{(1-T\Me)^2-T^2(1-\Me)^2/4}
\Pmatrix{1-T\Me&T(1-\Me)/2\\ T(1-\Me)/2&1-T\Me}.
\end{equation}
Expanding around $\Me=-1/3$ we have
\begin{equation}
\frac{1}{S}\sim\frac{1}{2\sqrt{3}\sqrt{1+3\Me}}\Pmatrix{1&-1\\-1&1}.
\label{eq:Laurent1/S}
\end{equation}
Therefore, the degree of singularity of $S^{-1}$ is one. Since
the twist-odd part $A$ has degree $-1$, every term in the infinite
series (\ref{eq:(S+A)^-1}) apparently has the same degree of
singularity and it might seem that we cannot simplify the expression
(\ref{eq:(S+A)^-1}) further.
This is, however, not true.
Let us first consider the term $S^{-1}AS^{-1}$.
The most singular part of $S^{-1}AS^{-1}$ is given by using
(\ref{eq:Laurent1/S}) as
\begin{equation}
\frac{1}{S}A\frac{1}{S}\sim
\frac{1}{2\sqrt{3}\sqrt{1+3\Me}}
\frac{\Mo}{2}\frac{1}{2\sqrt{3}\sqrt{1+3\Me}}
\Pmatrix{1&-1\\ -1&1}\Pmatrix{&1\\ -1&}\Pmatrix{1&-1\\ -1&1},
\end{equation}
which vanishes since the product of the three $2\times 2$ matrices is
equal to zero.
The less singular part in  $S^{-1}AS^{-1}$ does not contribute
terms with degree of singularity equal to three in $G$.
This argument applies to all the remaining terms in the expansion
(\ref{eq:(S+A)^-1}). Therefore, in the deformation of (\ref{eq:G}),
we are allowed to freely use the non-linear relations for
$(1-T\calM)^{-1}$ and hence use the expression (\ref{eq:(1-TM)^-1}).

Finally, let us comment on possible $T$ and $\bm{t}$ other than those
used in this paper.
We shall first comment on $\bm{t}$.
One might think that we should solve the original equation
(\ref{eq:u}) for $\bm{t}$ in the level truncation without using the
non-linear relations.
However, this is impossible if we impose the twist-even condition
(\ref{eq:Ct=t}) on $\bm{t}$.
In fact, eq.\ (\ref{eq:u}) and the one obtained from it by
multiplying $C$ and using (\ref{eq:Ct=t}) are over-determined for
$\bm{t}$. As explained in sec.\ 4.2 of \cite{HatKaw}, these equations
are consistently solved owing to the non-linear relations.
On the other hand, the original equation (\ref{eq:eqforT}) for $T$ in
the level truncation can have twist-even solutions without using the
non-linear relations. We do not know whether such solutions are
superior to the conventional one (\ref{eq:T-}) in any respects.

\section{Potential height problem revisited}
If the classical solution $\Psic$ of VSFT represents a D25-brane, the
energy density $\calEc$ of this solution relative to that of the
trivial one $\Psi=0$ should be equal to the D25-brane tension
$T_{25}$.
In \cite{HatKaw} they obtained the ratio $\calEc/T_{25}$ in a closed
form using the Neumann coefficients.
They further used the non-linear relations freely to simplify the
expression of $\calEc/T_{25}$ and calculated it numerically using
the level truncation. The result was not, however, the expected one.
Now we know that naive use of the non-linear relations is dangerous.
So we shall reexamine the ratio $\calEc/T_{25}$ by taking into account
the lesson we learned in the previous section.
We shall find that $\calEc/T_{25}$ is expressed in terms of a quantity
(called $H$ in \cite{HatKaw}) which, like $G$, vanishes if we freely
use the non-linear relations but give a non-vanishing value due to
twist anomaly.

Before reexamining the ratio $\calEc/T_{25}$, we shall first summarize
the derivation given in \cite{HatKaw}.
First, the energy density of the solution $\Psic$ is given by
\begin{equation}
\calEc=-\frac{\calS[\Psic]}{V_{26}}
=\frac{K}{6}\braket{\phic}{\phic}
=\frac{K}{6}
\left(\frac{\left[\det(1-T\calM)\right]^2}{\det(1-T^2)}\right)^{13}
\left(\frac{
[\det(1-\wt{T}\wt{\calM})]^2}{\det(1-\wt{T}^2)}\right)^{-1} ,
\label{eq:calEc}
\end{equation}
where we have used the equation of motion (\ref{eq:eqmot}) at the
second equality.
Next, let us calculate the D25-brane tension, which is given in the
present convention of $\ap=1$ by $T_{25}=1/(2\pi^2\go^2)$ with
$\go$ being the open string coupling constant defined as
the three-tachyon on-shell amplitude.
Using the tachyon wave function $\Phit$ (\ref{eq:phit}),
$\go$ is given by
\begin{align}
\go&=K\,\Phit\cdot(\Phit *\Phit)
\Bigr\vert_{p_1^2=p_2^2=p_3^2=-\mt^2}\nn\\
&=K\calNt^3\,\left[\det(1-T\calM_3)\right]^{-13}
\det(1-\wt{T}\wt{\calM}_3)
\,\exp\biggl\{
-\frac12\bm{V}(1-T\calM_3)^{-1}TC\bm{V}
\nn\\
&\qquad\qquad\qquad
+\bm{V}(1-T\calM_3)^{-1}\bm{t}A_0
-\frac12 A_0\bm{t}\calM_3(1-T\calM_3)^{-1}\bm{t}A_0
-\frac12 V_{00}(A_0)^2
\biggr\}.
\label{eq:KphitphitphitV}
\end{align}
Precisely speaking, we must remove
$(2\pi)^{26} \delta^{26}(\sum_{r}p_r)$ from the second term
$K\Phit\cdot(\Phit *\Phit)$.
The normalization factor $\calNt$ for $\Phit$ in (\ref{eq:phit})
is determined by the following requirement that
$\Phit$ has a canonical kinetic term:
\begin{equation}
\frac{K}{2}\Phit\cdot\calQB\Phit\underset{p^2\sim -\mt^2}{\sim}
-\frac12(p^2 +\mt^2),
\end{equation}
where we have omitted the momentum conservation delta function.
We have
\begin{align}
\calNt=\frac{1}{\sqrt{KG}}
\left[\det(1-T^2)\right]^{13/2}\bigl[\det(1-\wt{T}^2)\bigr]^{-1/2}
\exp\Bigl(\bm{t}\,(1+T)^{-1}\bm{t}\,\mt^2\Bigr).
\label{eq:calNt}
\end{align}
Collecting all these facts, we find the following expression for
the ratio $\calEc/T_{25}$:
\begin{align}
\frac{\calEc}{T_{25}}=\frac{\pi^2}{3G^3}\exp(6\mt^2H),
\label{eq:ratio}
\end{align}
with $H$ defined by
\begin{align}
H=&-\frac{2}{(A_0)^2}
\left[-\frac12\bm{V}(1-T\calM_3)^{-1}TC\bm{V}
+\bm{V}(1-T\calM_3)^{-1}\bm{t} A_0
-\frac12 A_0\bm{t}\calM_3(1-T\calM_3)^{-1}\bm{t} A_0
\right]\nn\\[7pt]
&\quad + \bm{t}(1+T)^{-1}\bm{t} + V_{00} .
\label{eq:totalH}
\end{align}
All the determinant factors in (\ref{eq:calEc}),
(\ref{eq:KphitphitphitV}) and (\ref{eq:calNt}) have been cancelled out
in (\ref{eq:ratio}) by use of the non-linear relations.

If the classical solution $\Psic$ represents a single D25-brane, the
value of $H$ must be equal to $H=(1/6)\ln(\pi^2/(3(\ln 2)^3))
\simeq 0.3817$ (we have used that $\mt^2=-1$ and hence $G=\ln 2$).
Similarly to $G$, we can show that $H$ vanishes by freely using the
non-linear relations, implying that $H$ is regarded as a twist anomaly
which needs careful treatments.
In \cite{HatKaw} they deformed $H$ (\ref{eq:totalH}) by using
the non-linear relations to obtain another expression
((5.13) of \cite{HatKaw}), which gave strange numerical
values.\footnote{
Using the general argument of sec.\ 3, we see that the value of $H$ in
the form of (5.13) of \cite{HatKaw} is equal to $-G/2$.
}
However, their manipulations contain forbidden ones in the sense of
sec.\ 3.
We have to reexamine $H$ by following the prescription of sec.\ 3.

To simplify the expression (\ref{eq:totalH}) for $H$ without changing
the ordering among the matrices, note first that $\calM_3$ is
partially diagonalized as follows:
\begin{align}
\calM_3=\UW^\dagger\Pmatrix{1&&\\ &\Up&\\ &&\Um}\UW ,
\end{align}
where $U_\pm$ and the unitary matrix $\UW$ are defined by
\begin{align}
U_\pm&=M_0+\omega^{\pm 1}M_++\omega^{\mp 1}M_- ,
\\
\UW&=\frac{1}{\sqrt{3}}
\Pmatrix{1&1&1\\ 1&\omega^2&\omega\\ 1&\omega&\omega^2},
\end{align}
with  $\omega=e^{2\pi i/3}$.
Using this basis and the fact that $\bm{V}$ in
(\ref{eq:newquantities}) is given in terms of $\ve$ and $\vo$ as
\begin{equation}
\bm{V} =
\frac{\vo}{2}
\Pmatrix{a^{(2)}_0-a^{(3)}_0\\a^{(3)}_0-a^{(1)}_0\\a^{(1)}_0-a^{(2)}_0}
+\frac{3\ve}{2}A_0 ,
\end{equation}
we obtain an expression of $H$ in terms of smaller matrices:
\begin{align}
H=&-\frac38
\vo\calUp T\vo -\frac{3\sqrt{3}i}{4}\vo\calUm T\ve
+\frac98\ve\,\calUp T\ve +\frac{\sqrt{3}i}{2}\vo\calUm\bm{t}
-\frac32\ve\,\calUp\bm{t}
\nn\\
&+\frac12\bm{t}\Bigl(\Up(1-T\Up)^{-1}
+\Um(1-T\Um)^{-1}\Bigr)\bm{t}+\bm{t}(1+T)^{-1}\bm{t} +V_{00},
\label{eq:newH}
\end{align}
with
\begin{equation}
{\cal U}^{\pm}=(1-T\Up)^{-1}\pm (1-T\Um)^{-1}.
\end{equation}
In deriving (\ref{eq:newH}), we have used following formula
\begin{align}
(a^{(1)}_0+\omega a^{(2)}_0+\omega^2a^{(3)}_0)
(a^{(1)}_0+\omega^2a^{(2)}_0+\omega a^{(3)}_0)
=9,
\end{align}
which comes from the on-shell condition $(a^{(r)})^2=-2\mt^2=2$.
Let us emphasize here again that the matrix ordering has been kept in
deriving (\ref{eq:newH}) from the original expression
(\ref{eq:totalH}).

We have calculated numerically the value of $H$ using the expression
(\ref{eq:newH}). The result is given in table \ref{tab:H}\,(a).
Amazingly, the ratio $\calEc/T_{25}$ seems to converge to $2$ in
contrast to our original expectation of $1$.
If there are no other subtle points in our analysis, this result
implies that the classical solution $\Psic$ represents the
configuration of two D25-branes.
\begin{table}[htbp]
\begin{center}
\parbox{5cm}{
\begin{tabular}[b]{|r|l|l|}
\hline
$L$~~ & ~~~$H$ & ~$\calEc/T_{25}$\\
\hline\hline
$50$ & $0.27398$ & $1.9089$ \\
$100$ & $0.27260$ & $1.9247$ \\
$150$ & $0.27179$ & $1.9341$ \\
$200$ & $0.27124$ & $1.9405$ \\
$250$ & $0.27083$ & $1.9453$ \\
$300$ & $0.27051$ & $1.9490$ \\
$400$ & $0.27003$ & $1.9546$ \\
$500$ & $0.26967$ & $1.9589$ \\
$600$ & $0.26939$ & $1.9622$ \\
$800$ & $0.26898$ & $1.9670$ \\
$1000$ & $0.26867$ & $1.9707$ \\
\hline
\end{tabular}
\centerline{(a)}
}
\hspace{2cm}
\parbox{5cm}{
\begin{tabular}[b]{|r|l|l|}
\hline
$L$~~ & ~~~$H$ & ~$\calEc/T_{25}$\\
\hline\hline
$50$ & $0.25667$ & $2.1178$ \\
$100$ & $0.26119$ & $2.0611$ \\
$150$ & $0.26269$ & $2.0427$ \\
$200$ & $0.26341$ & $2.0338$ \\
$250$ & $0.26383$ & $2.0287$ \\
$300$ & $0.26409$ & $2.0256$ \\
$400$ & $0.26440$ & $2.0218$ \\
$500$ & $0.26456$ & $2.0199$ \\
$600$ & $0.26465$ & $2.0188$ \\
$800$ & $0.26473$ & $2.0178$ \\
$1000$ & $0.26476$ & $2.0174$ \\
\hline
\end{tabular}
\centerline{(b)}
}
\caption{The values of $H$ in the level truncation calculation.
In table (a), we used the expression (\ref{eq:newH}), while in (b)
we used (\ref{eq:simpleH}).
For $\calEc/T_{25}$ we used (\ref{eq:ratio}) with $G=\ln 2$ and
$\mt^2=-1$.
}
\label{tab:H}
\end{center}
\end{table}

Following the prescription explained in sec.\ 3, $H$ given by
(\ref{eq:newH}) can safely be deformed as follows.
Taylor-expanding $1-TU_\pm$ around $M_0=-1/3$, we have
\begin{equation}
1-TU_\pm\sim\sqrt{3}\sqrt{1+3M_0}\pm\frac{\sqrt{3}i}{2}\Mo .
\end{equation}
Using this, we obtain a simpler expression of $H$:
\begin{align}
&H=\frac{\sqrt{3}}{4}\vo\frac{1}{\sqrt{1+3\Me}}\biginv\vo
-\frac{3\sqrt{3}}{8}\vo\frac{1}{\sqrt{1+3\Me}}\Mo
\frac{1}{\sqrt{1+3\Me}}\biginv\frac{1}{\sqrt{1+3\Me}}\Mo\vo\nn\\
&-\frac{9\sqrt{3}}{16}\vo\Mo\frac{1}{1+3\Me}\biginv
\frac{1}{\sqrt{1+3\Me}}\Mo\vo
+\frac{9\sqrt{3}}{16}\vo\Mo\frac{1}{(1+3\Me)^{3/2}}\Mo\vo
+H_{\rm reg},
\label{eq:simpleH}
\end{align}
where $\biginv$ is defined by
\begin{align}
\biginv =\biggl(
1+\frac14\Mo\frac{1}{\sqrt{1+3\Me}}\Mo\frac{1}{\sqrt{1+3\Me}}
\biggr)^{-1} ,
\end{align}
and  $H_{\rm reg}$ is the part with degree of singularity less than
three. As we did for $\Greg$ in (\ref{eq:Gdeform}), $H_{\rm reg}$ is
given as the negative of the terms on the RHS of
(\ref{eq:simpleH}) other than $H_{\rm reg}$ calculated by using
naively the non-linear relations and expressed without $\Mo$.
Explicitly, we have
\begin{equation}
H_{\rm reg}= -\frac{\sqrt{3}}{16}\vo\frac{(1+3\Me)^{3/2}}{5-\Me}\vo .
\end{equation}
This simpler form (\ref{eq:simpleH}) of $H$ should give the
same value as the original one. In fact, the result of level
truncation calculation presented in table \ref{tab:H}\,(b) confirms
this expectation.

\section{Summary and future problems}

In this paper we have presented an interpretation as twist anomaly to
$G$ and $H$ expressing the physical observables in VSFT expanded
around a classical solution. We have reexamined the potential height
problem for the solution and obtained a result indicating that it
represents the configuration of two D25-branes.
Let us finish this paper by presenting future problems.
\begin{itemize}
\item
We have obtained in this paper a numerical result that
$\calEc/T_{25}=2$.
Though this result is not obviously strange,
it is not a natural one. In arriving at the formula (\ref{eq:ratio})
for the ratio $\calEc/T_{25}$, all the determinant factors in
(\ref{eq:calEc}) and (\ref{eq:calNt}) have been cancelled out among
them.
However, since the eigenvalue distribution of $\Me$ and $T$ extends to
$-1/3$ and $-1$, respectively, the cancellation of the determinants
contains indefinite quantities like $0/0$. In fact, numerical analysis
of the determinants given in \cite{HatKaw} indicates that the
cancellation among the determinants is subtle.
We have to clarify these points for obtaining the final answer to the
ratio $\calEc/T_{25}$.

\item
In this paper, we have calculated $G$ and $H$ numerically by using
the level truncation. It is of course desirable to develop a method to
calculate them analytically. For example, exact expression of
eigenvalue distribution function for $\Me$ would be a first step
toward this subject.

\item
The matrix $T$ used in this
paper has been obtained by freely using the non-linear relations
among $M_\alpha$ on the original equations for $T$.
As mentioned at the end of sec.\ 3, there are other candidate
solutions for $T$ obtained without using the non-linear relations.
We have to examine whether such solutions give different results for
physical quantities.

\item
The most important and interesting problem for VSFT is to show that
the perturbation theory expanded around the trivial configuration
$\Psi=0$  reproduces pure closed string theory.
As seen in this paper, physical observables in VSFT are interpretable
as twist anomaly. Closed string might also emerge as a twist anomaly.
\end{itemize}

\section*{Acknowledgments}
We would like to thank S.\ Dobashi and I.\ Kishimoto for valuable
discussions and comments.
The works of H.\,H.\ and S.\,M.\ were supported in part by a
Grant-in-Aid for Scientific Research from Ministry of Education,
Culture, Sports, Science, and Technology (\#12640264 and \#04633,
respectively).
S.\,M.\ is supported in part by the Japan Society for
the Promotion of Science under the Predoctoral Research Program.

\end{document}